\documentclass[runningheads]{lncse}

\usepackage{amsmath,amssymb,amsfonts,latexsym}
\usepackage{changebar}
\usepackage{epsfig}

\begin{document}

\mainmatter    

\newcommand{\hpd}{{\em hpd}}
\newcommand{\uq}{{\sf u}}
\newcommand{\dq}{{\sf d}}
\newcommand{\sq}{{\sf s}}
\newcommand{\cq}{{\sf c}}
\newcommand{\bq}{{\sf b}}
\newcommand{\tq}{{\sf t}}
\newcommand{\ie}{{\mbox{\frenchspacing{i.\hspace{0.04cm}e.}}}}
\newcommand{\etc}{{\frenchspacing{etc.}}}
\newcommand{\nn}{\nonumber}
\newcommand{\buno}{{\mathbf 1}}

\title{
{\rm\small \hspace*{\fill}WUB 00/16}\\
One-Flavour Hybrid Monte Carlo with Wilson Fermions}

\titlerunning{One-Flavour Hybrid Monte Carlo with Wilson Fermions}

\author{Thomas Lippert}
\authorrunning{Th.\ Lippert}

\institute{University of Wuppertal\\ Department of Physics\\
D-42097 Wuppertal, Germany}

\maketitle

\begin{abstract}
  The Wilson fermion determinant can be written as product of the
  determinants of two hermitian positive definite matrices.  This
  formulation allows to simulate non-degenerate quark flavors by means
  of the hybrid Monte Carlo algorithm. A major numerical difficulty is
  the occurrence of nested inversions.  We construct a Uzawa
  iteration scheme which treats the nested system within one iterative
  process.
\end{abstract}

\index{determinant!fermionic}
\index{Wilson fermions}
\index{matrix!hermitian positive definite}

\label{PTHOMAS}

\section{Introduction}

\index{QCD}
\index{QCD!hadronic mass scale}
\index{quantum chromodynamics}
\index{QCD!on the lattice}
\index{lattice!QCD}
\index{lattice!gauge theory}
\index{lattice!regularization}
\index{strong interaction}
\index{standard model}
\index{particle physics}

The strong interaction between the quarks is described by quantum
chromodynamics (QCD), a constitutive part of the standard model of
elementary particle physics.  QCD develops a strong coupling constant
at the hadronic mass scale and thus, perturbation theory cannot be
applied. The only known non-perturbative ab-initio method is to
simulate QCD on a 4-dimensional space-time lattice by use of Monte
Carlo methods as known from statistical physics.

\index{Monte Carlo simulation}
\index{statistical physics}
\index{closed fermion loops}
\index{space-time!four-dimensional}
\index{ab-initio simulation}
\index{quark!flavors}
\index{quark-antiquark pair}
\index{chiral symmetry!dynamical breaking of}

One would think that lattice simulations of QCD have to take into
account virtual loops from all six quark flavors, up (\uq), down
(\dq), strange (\sq), charm (\cq) bottom (\bq), and top (\tq).  Their
masses span a wide range from about $3$ MeV to about $180$ GeV.
\footnote{The light quark (\uq, \dq, \sq) masses are so-called
  ``current'' masses determined within the $\overline{\mbox{MS}}$
  scheme at a renormalization scale $\mu=2$ GeV. The masses of the
  heavy quarks (\cq, \bq, \tq) are ``running'' masses determined at
  $\mu=m_{\mbox{\footnotesize\cq,\bq,\tq}}$ in the
  $\overline{\mbox{MS}}$ scheme \cite{Caso:1999pd}.}  The masses of
the three heavy quarks lie above the momentum scale set by dynamical
chiral symmetry breaking.  The QCD Lagrangian is chirally symmetric
for vanishing quark masses.  Chiral symmetry is explicitly broken by
the quark masses and spontaneously by the dynamics.  For light quarks,
(\uq, \dq, \sq), dynamical breaking dominates, as signaled by zero
mass Goldstone bosons, \ie, the pion octet.  Supposedly only the light
quarks contribute with virtual loops; the contributions of the heavier
ones are assumed to be negligible.

\index{Goldstone boson}
\index{pion octet}
\index{light quarks}
\index{light quarks!small mass of}
\index{chiral perturbation theory}
\index{correlation length}
\index{msbar@{$\overline{\mbox{MS}}$ scheme}}

A straightforward lattice QCD simulation including \uq, \dq, and \sq\ 
is difficult: on one hand, \uq\ and \dq\ are very light.  As we know
from chiral perturbation theory, the square of the pion mass is
proportional to the quark mass \cite{Gasser:1982ap}.  Therefore, the
pion acquires a small mass, \ie\ a large correlation length,
$\xi=\frac{1}{ma}$, for \uq\ and \dq\ masses approaching the chiral
limit, $m_{\mbox{\footnotesize \uq}}$ and $m_{\mbox{\footnotesize
    \dq}}=0$.  $a$ denotes the lattice spacing. The lattice volume
must be larger by more than one order of magnitude than possible today
to reach physical parameter values for $m_{\mbox{\footnotesize \uq}}$
and $m_{\mbox{\footnotesize \dq}}$.  Hence, one has to recourse to
extrapolations using results of simulations at artificial parameters
for $m_{\mbox{\footnotesize \uq}}$ and $m_{\mbox{\footnotesize \dq}}$
and small lattices, far off the chiral limit. These difficulties go
along with an increasing condition number of the fermionic matrix $M$,
therefore its inversion suffers from critical slowing down approaching
the chiral limit \cite{Frommer:1994vn,Fischer:1996th}.

\index{chiral limit}
\index{lattice!volume}
\index{condition number} 
\index{Dirac operator!condition number of}
\index{critical slowing down}
\index{stochastic!estimate}
\index{hybrid Monte Carlo!exact algorithm}
\index{HMC}

On the other hand, until recently, the only exact simulation algorithm
for QCD with dynamical fermions\footnote{``Dynamical'' in contrast to
  ``quenched'' simulations that neglect fermionic loops.} was the
hybrid Monte Carlo algorithm (HMC)
\cite{Duane:1987de,Gottlieb:1987mq}. The benefits of HMC in reducing
the computational complexity are achieved by treating the fermionic
determinants as a stochastic estimate using Gaussian random fields
\cite{Weingarten:1981hx}, with the advantage, that instead of
computing the determinant, only the solution of a linear system is
required.  For this approach the fermion matrix $M$ must be hermitian
positive definite (\hpd\/).

\index{matrix!hermitian positive definite}
\index{Wilson fermions}
\index{QCD!one flavor}
\index{QCD!single Wilson flavor}
\index{one flavor QCD}
\index{fermion doublers}
\index{doubling problem}
\index{chiral symmetry!explicit violation of}
\index{matrix!complex}
\index{matrix!non-normal}

The Wilson fermion discretization of the fermionic sector of QCD
describes single flavors \cite{Wilson:1975hf}, in contrast to
staggered fermions that represent four fermions, intermixed in
spin-flavor space.\footnote{The naive discretization of the Dirac
  operator yields, in addition to the true mode, 15 ``doublers''.}
Unfortunately, Wilson fermions must violate chiral symmetry, see
Ref.~\cite{Neuberger:1999zk}. As a consequence, the Wilson fermion
matrix is complex non-normal and thus cannot be included in the HMC
scheme. To avoid this problem, one usually simulates two mass
degenerate light quarks, an approximation justified by the fact that
both quarks, \uq\ and \dq, are light and close in mass compared to the
next heavier one, \sq.  The product of two mass degenerate
determinants indeed amounts to an \hpd\ matrix, that can be simulated
by the HMC.  The minor price to pay is that the two light quarks are
mass-degenerate, the major price is that the \sq\ quark is not
included in the simulation.

\index{light quarks!mass degeneracy}

There are attempts to evaluate operators, that contain the \sq\ quark,
by application of a ``partially quenched analysis''
\cite{Sharpe:1999kj}.  The PQA extrapolates only one of the valence
quarks towards the chiral limit and holds the other one at the mass of
the \sq\ quark---of course within the light mass-degenerate
\uq-\dq-sea. PQA, however, leads to inconsistencies: the so-called
J-parameter does not acquire its physical value
\cite{Eicker:1998sy,AliKhan:1999zp}.

\index{partially quenched analysis}
\index{quenched!approximation}
\index{quenched!partially}
\index{valence quark}
\index{J-parameter}

One can try to employ an approximate one-flavor determinant
representation. Such simulations, however, are plagued by
systematic uncertainties, as one has to recourse to non-exact
simulation algorithms like the hybrid molecular dynamics algorithm
\cite{Bernard:2000gd,Peikert:1998jz}.

\index{hybrid molecular dynamics}
\index{hybrid molecular dynamics!non-exact algorithm}

On the other hand, in the framework of the multi-boson algorithm
\cite{Luscher:1994xx}, one can define an exact single flavor
simulation scheme for Wilson fermions
\cite{Alexandrou:1998wv,Alexandrou:1998pb,Montvay:1999tp}.  Here many
additional bosonic auxiliary fields are required, and it is not clear
if the method would be well suited for more complicated actions or
allows to exploit sophisticated preconditioning.

\index{multi-bosonic algorithm}
\index{multi-bosonic algorithm!two step}

In this paper, we propose to include the Wilson fermion determinant
into HMC such that a single quark can be described.  We show that the
fermionic determinant can be written as a product of the determinants
of two \hpd\ matrices.  This is achieved by the representation of the
fermion matrix via its Schur complement.  Both determinants can be
treated by the hybrid Monte Carlo in the standard manner. There are,
however, some caveats: one of the two matrices involves the inverse of
the other one. As a consequence, a nested inversion must be carried
out. We propose to use a Uzawa-like inverter which allows to solve the
nested system within one iterative process.  A second difficulty stems
from the initialization of the fermion action in the HMC algorithm:
since we have a single flavor system, we have to compute a square root
at the beginning of a HMC trajectory. 

\index{Schur complement}
\index{iterative solver!nested}
\index{nested inversion}
\index{Uzawa iteration}
\index{square root of matrix}
\index{matrix!square root of}

The paper is organized as follows: in section \ref{SEC:NOTATIONS} we
introduce the basics of lattice QCD and will discuss the difficulties
simulating single flavors of Wilson fermions by the HMC. In section
\ref{SEC:TRANSFORM} we give the transformation of the fermion matrix
using the Schur complement, in section \ref{SEC:HMC} we present the
non-mass-degenerate HMC for Wilson fermions along with a discussion of
numerical problems.  In section \ref{SEC:UZAWA}, we introduce an
algorithm that can compute the nested inversion in one iterative
process. Finally, we summarize and give an outlook.

\section{Numerical Problem\label{SEC:NOTATIONS}}

\index{QCD!action}
\index{QCD}
\index{QCD!on the lattice}
\index{lattice!QCD}
\index{lattice!gauge theory}
\index{lattice!regularization}
\index{quantum chromodynamics}
\index{Monte Carlo simulation}
\index{Euclidean!space}
\index{space-time!four-dimensional}

The action of lattice QCD is the sum of a gauge part (which is not
relevant for the following) and a fermionic part:
\begin{equation}%
S[U,\bar\psi,\psi]:=S_g[U]+\sum_{{\sf x},{\sf y},a,b,\alpha,\beta}\bar\psi^{a\atop\alpha}_{\sf x}
\,M^{a,b\atop \alpha,\beta}_{{\sf x},\sf y}\,\psi^{b\atop \beta}_{\sf y}.
\end{equation}%
$V$ is 4-dimensional volume in Euclidean space.  The full matrix is a
tensor product with $3\times 3$ SU(3) matrices $U$ (color) and four
$4\times 4$ matrices $\gamma_{\mu}$ (Dirac spin), hence
\begin{equation}%
M \in {\mathbb C}^{12V\times 12V}.
\end{equation}%
Here the indices $\sf x$ an $\sf y$ stands for the 4-dimensional
coordinates, $a$ and $b$ denote color space and $\alpha$ and $\beta$
Dirac space indices.

\index{unitary matrix}
\index{link element}
\index{SU(3)}
\index{index!colour}
\index{index!Dirac}
\index{index!space-time}
\index{Grassmann variables}
\index{gammamat@{$\gamma$ matrices}}
\index{Dirac matrices}
\index{gammafive@{$\gamma_5$}}
\index{path integral}

The fermion fields $\psi_{\sf x}$ are {\em Grassmann} variables.
Therefore, the Grassmann integral of the exponential of the fermionic
part of the action, \ie, the fermionic part of the path integral, can be
carried out,
\begin{equation}%
\int [d\psi][d \bar\psi]\,e^{-\bar\psi\,M\psi}=\det(M),
\end{equation}%
to yield the determinant of $M$ for arbitrary complex $M$.

Wilson fermions are defined by the interaction matrix
\begin{eqnarray}
M &=& \buno - \kappa D,\qquad M \in {\mathbb C}^{12V\times 12V},
\nn\\
D_{\sf x,y} &=& \sum_{\mu=1}^4 
{(\buno-\gamma_{\mu})}\,U_{\mu}(\sf x)\,\delta_{\sf x,\sf y-\mu}+
{(\buno+\gamma_{\mu})}\,U_{\mu}^{\dagger}(\sf x-\mu)\,\delta_{\sf x,\sf y+\mu}.
\end{eqnarray}

The Dirac matrices in Euclidean space satisfy the anti-commutator relations
\begin{equation}%
\gamma_{\mu}
\gamma_{\nu}
+
\gamma_{\nu}
\gamma_{\mu}
=
2\delta_{\mu\nu}.
\end{equation}%
It is important for the following considerations to work with the
chiral representation of the Dirac matrices, as given in the appendix
\ref{CHIRALAPP}.  In this representation, $\gamma_5$, defined as the
product $\gamma_5 = \gamma_4\gamma_1\gamma_2\gamma_3$, is diagonal.

\index{Wilson fermions}
\index{gammamat@{$\gamma$ matrices}!anti-commutator relations of}
\index{gammamat@{$\gamma$ matrices}!chiral representation}
\index{matrix!complex}
\index{matrix!non-normal}
\index{similarity transform}
\index{gammafive@{$\gamma_5$}!hermiticity}
\index{Dirac operator!gammafive@{$\gamma_5$} hermiticity}

$M$ is a complex non-hermitian matrix which moreover is not normal, \ie,
\begin{equation}%
MM^{\dagger}\ne
M^{\dagger}M.
\end{equation}%
Thus $M$ cannot be diagonalized by a unitary transformation.  At most,
it might be diagonalizable by a similarity transformation.  However,
$M$ exhibits the so-called $\gamma_5$-symmetry or
$\gamma_5$-hermiticity:
\begin{equation}%
M\gamma_5 = \gamma_5M.
\end{equation}%
Therefore, the matrix
\begin{equation}%
Q:=\gamma_5M
\end{equation}%
is hermitian.  We can immediately read off the chiral representation
of the Dirac matrices, see the appendix, that, for $\kappa=0$, $Q$
exhibits an equal number of positive and negative eigenvalues. In
general, $Q$ is indefinite.
\index{eigenvalue!equal number of positive and negative}
\index{matrix!indefinite}

The matrix $M^{\dagger}M$ is \hpd. Since the determinant of
$M^{\dagger}M$ represents two mass degenerate fermion flavors,
\begin{equation}%
\det\big(M^{\dagger}M\big) = 
\det\big(M\big)  
\det\big(M\big),  
\end{equation}%
the two flavor situation is ideal for HMC simulations. Being unitary
diagonalizable it can be represented by a Gaussian integral over
complex fields $\phi$:

\index{path integral!Gaussian}
\index{Gaussian!integral}

\begin{equation}%
\det\big(M\big)  
\det\big(M\big)= \left(\frac{1}{2\pi}\right)^{12V}
\int [d\phi][d\phi^*]
e^{-\phi^{\dagger}\big(M^{\dagger}M\big)^{-1}\phi}.
\end{equation}%

For the single flavor $M$, such a simple construction fails even for
the situation $\kappa < \kappa_c$ where $M$ is positive definite,
because the Jacobian of the similarity transform---if the
diagonalization is feasible at all---is not equal to 1.  

On the other hand,
\begin{equation}%
\det\big( M\big)=
\det\big( Q\big).
\end{equation}%
But, $Q$ is indefinite. Therefore, a Gaussian integral representation
for its determinant does not exist.

We conclude that so far there is no direct way to include individual
flavors within the HMC algorithm.

\section{Schur Complement of $Q$\label{SEC:TRANSFORM}}

Let's consider the hermitian matrix $Q$ with the Dirac matrices given
in the chiral representation of appendix \ref{CHIRALAPP}.  The
explicit form of $Q$ in the chiral representation is given by
\begin{equation}%
Q=\left(
\begin{array}{ll}
Q_{11} &
Q_{12} \\
Q_{21} &
Q_{22} \\
\end{array}
\right)
=\left(
\begin{array}{cc}
{\mathbf 1}
(1-\kappa D_{11}) &
-\kappa D_{12}  \\
-\kappa D^{\dagger}_{12}  &
-{\mathbf 1} (1-\kappa D_{11})
\\
\end{array}
\right).
\end{equation}%
Here, the bold face ${\mathbf 1}$ represents a unit matrix in $2\times
2$ spin space.  $D_{11}$ does not carry spin indices while $D_{12}$
consists of $2\times 2$ blocks in spin space,
\begin{eqnarray}%
D_{11} &=& \sum_{\mu=1}^4 U_{\mu}(\sf x)\,\delta_{\sf x,\sf x+\mu}+
U_{\mu}^{\dagger}(\sf x-\mu)\,\delta_{\sf x,\sf x-\mu}\nn\\
D_{12} &=& \sum_{\mu=1}^4 \eta_{\mu}\, \big[U_{\mu}(\sf x)\,\delta_{\sf
  x,\sf x+\mu}- U_{\mu}^{\dagger}(\sf x-\mu)\,\delta_{\sf x,\sf
  x-\mu}\big],
\label{DIAGSPIN}
\end{eqnarray}
with
\begin{equation}%
\begin{array}{cc}
\eta_{1}=i\sigma_1,  & \eta_{2}=i\sigma_2, \\
\eta_{3}=i\sigma_3, & \eta_{4}=-{\mathbf 1}.
\end{array}
\end{equation}%

The Schur complement of a $2\times 2$ block matrix with non-singular
block $A$ is given by
\begin{equation}%
\left[\begin{array}{cc}
A & B \\
C & D \\
\end{array}\right]
=
\left[\begin{array}{cc}
{\mathbf 1} & 0 \\
CA^{-1} & {\mathbf 1} \\
\end{array}\right]
\left[\begin{array}{cc}
A & 0 \\
0 & D-C\,A^{-1}\,B \\
\end{array}\right]
\left[\begin{array}{cc}
{\mathbf 1} & A^{-1}B \\
0 & {\mathbf 1} \\
\end{array}\right].
\end{equation}%
Hence
\begin{equation}%
\det(Q)  =\det\Big({\mathbf 1}(1-\kappa D_{11})\Big)\det\left(
{\mathbf 1}(1-\kappa D_{11}) + \kappa^2 D_{12}
[{\mathbf 1}(1-\kappa D_{11})]^{-1}D_{12}^{\dagger}\right).
\end{equation}%
The minus sign of the $Q_{22}$ term cancels as the rank of the matrix
is even.  From Eq.~\ref{DIAGSPIN} we know that the first matrix is
diagonal in Dirac space. Therefore its determinant can be written as
follows:
\begin{equation}%
\det(Q)  =\det\big({\mathbf 1}(1-\kappa D_{11})\big) = 
[\det(1-\kappa D_{11})]^2 =  
\det(1-\kappa  D_{11})^2,   
\end{equation}%
where $Q_w^2=(1-D_{11})^2$ is \hpd\ for any $\kappa$. $Q_w$
carries no spin index.

The second matrix 
\begin{equation}%
Q_{sc}=
{\mathbf 1}(1-\kappa D_{11}) + \kappa^2 D_{12}
[{\mathbf 1}(1-\kappa D_{11})]^{-1}D_{12}^{\dagger},
\end{equation}%
the Schur complement, is hermitian, as it is the sum of hermitian
matrices. For $\kappa=0$, it is \hpd, thus there exists some value of
$\kappa$, $\kappa^{sc}_c$,
for which it becomes indefinite. For $\kappa<\kappa^{sc}_c$, $Q_{sc}$ is
\hpd.
\index{critical $\kappa$ value}

\section{One Flavor HMC\label{SEC:HMC}}

We have shown that
\begin{equation}%
\det\big(M\big) = 
\det\big(Q\big) = 
\det\big(Q_w\big)
\det\big(Q_{sc}\big),
\end{equation}%
for $Q_{11}$ being non-singular.  If $\kappa<\kappa_{sc}$, both
matrices are \hpd\ and can be represented by Gaussian integrals.  Let
$\phi\in{\mathbb C}^{3V}$ and $\chi\in{\mathbb C}^{6V}$:
\index{Gaussian!integral}
\begin{eqnarray}
\lefteqn{\det\big(Q_w\big)
\det\big(Q_{sc}\big)}\nn\\
&=& \left(\frac{1}{2\pi}\right)^{3V}\left(\frac{1}{2\pi}\right)^{6V}
\int
[d\phi][d\phi^*]
[d\chi][d\chi^*]
e^{
-\phi^{\dagger}\left(Q_w\right)^{-2}\phi
-\chi^{\dagger}\left(Q_{sc}\right)^{-1}\chi}.\nn\\
\end{eqnarray}

\index{pseudo fermion}

The fermionic part of the path integral being expressed in terms of
the pseudo-fermion degrees of freedom $\phi$ and $\chi$, the HMC
algorithm can proceed as usual \cite{Lippert:1997mo}, see the
$\Phi$-algorithm in Ref.~\cite{Gottlieb:1987mq}.  The heat bath for
the $\phi$ fields is trivial like in the previous case of
mass-degenerate QCD.  Let $R\in{\mathbb C}^{3V}$ be Gaussian
distributed. With
\begin{equation}%
\phi=Q_w^{\dagger}R,
\end{equation}%
\begin{equation}%
R^{\dagger}R=\phi^{\dagger}(Q_w)^{-2}\phi.
\end{equation}%
and 
\begin{equation}%
(2\pi)^{3V}
\det(Q_w^2) =\int
[d\phi][d\phi^{*}]
e^{- \phi^{\dagger}(Q_w)^{-2}\phi}.
\end{equation}%
The refreshment of the $\chi$ fields causes more trouble, however.  In
order to generate the required distribution, we have to solve a linear
system involving the square root of $Q_{sc}$.  Let $R\in{\mathbb
  C}^{6V}$ be Gaussian distributed:
\index{matrix!square root of}
\begin{equation}%
\chi={Q_{sc}}^{\frac{1}{2}} R.
\end{equation}%
Then 
\begin{equation}%
R^{\dagger}R=
\chi^{\dagger}(Q_{sc})^{-1}\chi
\label{RR}
\end{equation}%
and
\begin{equation}%
(2\pi)^{6V}
\det(Q_{sc}) =\int
[d\chi][d \chi^{*}]
e^{-\chi^{\dagger}(Q_{sc})^{-1}\chi}
\end{equation}%
Even though the solution of Eq.~\ref{RR} is only required at the beginning
of a trajectory, it is particularly expensive as we must cope with the
internal inversion of $Q_{11}$. The implications of this issue are not
clarified yet.  As an alternative, we might include the dynamics of
the $\chi$ fields into HMC avoiding the expensive heat bath step for
$Q_{sc}$, similar to  the $R$-algorithm of Ref.~\cite{Gottlieb:1987mq}.

\index{heat bath}

\section{Inexact Uzawa Iterations\label{SEC:UZAWA}}
\index{iterative solver!nested}
\index{iterative solver!nested}

The nested inversion within the solution of
\begin{equation}%
Q_{sc}X=\chi
\end{equation}%
\index{Uzawa iteration} has to be carried out at each molecular
dynamics time step of HMC as well as in the computation of the action.
By Uzawa iterations \cite{Bramble:1997uz} we avoid the solution of the
linear system represented by $Q_w$ within each iteration step of the
$Q_{sc}$ iteration.

The problem to solve is given by:
\begin{eqnarray}%
(Q_w+D_{12}\, Q_w^{-1}\, D_{12}^{\dagger})
\, X&=&\chi,
\label{QSC}
\end{eqnarray}
with $ X,\psi\in{\mathbb C}^{6V}$.

Let's consider the Jacobi iteration for $Q_{sc}$,
\begin{eqnarray}
 X^{(i+1)} &=& \chi
+\kappa D_{11} X^{(i)}
-D_{12}\,Q_w^{-1}\,D_{12}^{\dagger}
 X^{(i)}.
\label{JAC}
\end{eqnarray}
\index{fixed-point}
The fixed-point solution $X$ of Eq.~\ref{JAC}, defined by
\begin{equation}%
|| X^{(i+1)}- X^{(i)}||<\epsilon,\quad\epsilon\rightarrow 0
\end{equation}%
solves equation Eq.~\ref{QSC}, as can be easily verified.

Next we transform the simple Jacobi iteration into the Uzawa iteration:
\index{Jacobi!iteration}
\begin{eqnarray}%
 X^{(i+1)} &=& \chi
+\kappa D_{11} X^{(i)}
-D_{12}\,
 Y^{(i)}\nn\\
 Y^{(i+1)}&=&
Q_w^{-1}\,D_{12}^{\dagger}
 X^{(i)}.
\label{UZ}
\end{eqnarray}%
At this stage, the Uzawa scheme would certainly not offer an advantage
as still we need to carry out an inversion in each iteration step.
Let's turn towards the inexact Uzawa iteration.  Choose
\index{Uzawa iteration!inexact}
\index{inexact Uzawa iteration}
\begin{equation}%
P_w \approx Q_w
\end{equation}%
as a preconditioner for $Q_w$ which is easily invertible.
\begin{eqnarray}%
 X^{(i+1)} &=& \chi
+\kappa D_{11} X^{(i)}
-D_{12}\,
 Y^{(i)}\nn\\
 Y^{(i+1)}&=& Y^{(i)} +
P_w^{-1}(D_{12}^{\dagger}
 X^{(i)}-Q_w\,
 Y^{(i)})
\label{IUZ}
\end{eqnarray}
The fixed-point solution of Eq.~\ref{IUZ}, defined by
\begin{eqnarray}%
|| X^{(i+1)}- X^{(i)}||&<&\epsilon,\nn\\
|| Y^{(i+1)}- Y^{(i)}||&<&\epsilon'.
\end{eqnarray}%
for $\epsilon,\epsilon'\rightarrow 0$, solves Eq.~\ref{QSC}.  Using
the second relation, one finds
\begin{eqnarray}
0 &=&
P_w^{-1}(-Q_w\, Y+D_{12}^{\dagger}\, X),\nn\\
\Leftrightarrow
 Y&=& Q_w^{-1}D_{12}^{\dagger}\,  X.
\end{eqnarray}
As a cheap preconditioner we can use a truncated Neumann series for
$Q_w^{-1}$:
\index{Neumann series!truncated}
\begin{equation}%
P_w^{-1}=\big(1-\kappa\,D_{11}\big)^{-1}=1+\kappa\,D_{11}+\dots
\end{equation}%
One has to find the optimal length of this series compared to the
convergence of the outer iteration. In our implementation we have
chosen the simple first-order approximation. 
\begin{equation}%
P_w^{-1}=1+\kappa\,D_{11}.
\end{equation}%

\index{conjugate gradient}
\index{quenched!configuration}
\index{Fourier transform!diagonalization by}
\index{Krylov subspace}

We have made first tests of the Uzawa iteration in comparison to a
conjugate gradient (CG) for the inversion of Eq.~\ref{QSC}. We used a
quenched $16^4$ configuration at $\beta=6.0$ at a moderate
$\kappa$-value of $\kappa=0.12$. The outer CG took about 240
iterations while the inner iteration took about 50 steps.  The Uzawa
iteration required about 610 steps in comparison. The overall
improvement for this setting is about a factor of 10. Of course, these
results are very preliminary and have to be confirmed for other values
of $\kappa$ and $\beta$ with realistic HMC simulations.  It would also
be desirable to implement the Uzawa scheme within Krylov subspace
solvers.

\index{SESAM collaboration}

For free theory ($U=1$), one can diagonalize $D_{11}$ by Fourier
transformation.  As in the case of Wilson fermions, one finds
$\kappa_c=\frac{1}{8}$. We have, so far, investigated the lowest lying
eigenvalues of $Q_w$ for the interacting case: on a $16^3\times 32$
configuration, taken from the SESAM ensembles at $\beta=5.6$ and
$\kappa_{\mbox{\footnotesize sea}}=0.1575$, we found for the inverse
of the critical eigenvalue of $D_{11}$, $1/\lambda_{min}=0.144$.  The
critical $\kappa$ of $D_{11}$ is reached for a $\kappa$-value being
smaller than the critical $\kappa$ value of the Wilson matrix.
Therefore, the singularity might be a barrier screening the approach
to the chiral limit.  These questions have to be clarified in an actual
HMC simulation with one-flavor QCD. 

\section{Summary and Outlook}

The determinant of the Wilson fermion matrix is equivalent to the
product of the determinants of two hermitian positive definite
matrices. This formulation allows to simulate quark flavors with
individual mass values by means of the HMC algorithm. The solution of
the ensuing linear system is hampered by a nested inversion problem.
However, it can be treated by use of Uzawa iterations.

Currently we perform a feasibility study to gain first experiences
with the new method. In particular, we are interested in the analytic
structure of $Q_w$ and $Q_{sc}$. Furthermore, we construct an
efficient heat-bath for the refreshment of the pseudo-fermions for
$Q_{sc}$  based on the polynomial representation of the square
root of a truncated polynomial that represents $Q_{sc}$.

We have presented the method for standard Wilson fermions. The Schur
decomposition can as well be applied to improved Wilson fermions with
clover term. Work on this line is in progress.

\section*{Acknowledgments}
I thank B.\ Bunk for important discussions that motivated the present
work.  I am happy to work with A.\ Frommer, B\ Medeke, and K.\ 
Schilling from Wuppertal university who contributed with interesting
ideas.  In particular I thank H.\ Neff who has kindly performed the
eigenvalue computations. I am grateful to M.\ Peardon and Ph.\ de
Forcrand who pointed out important simplifications of the method
during my talk at the workshop.

\newpage
\appendix

\section{Chiral Dirac Matrices\label{CHIRALAPP}}

Euclidean Dirac matrices in the chiral representation:
\begin{equation}
\begin{array}{llcllr}
  \gamma_1 & = & \left(
\begin{array}{llll} 0 & 0 & 0 & -\mbox{i}  \\ 
         0  & 0 &     -\mbox{i} & 0 \\ 0 & \mbox{i} & 0 & 0 \\
      \mbox{i} & 0 & 0 & 0 \\
\end{array}\right) 
&=& \mbox{i}&\left(
\begin{array}{llll}  0 & -\sigma_1 \\ \sigma_1 & 0
\end{array}\right)\\
  \gamma_2 &=& 
\left(
\begin{array}{llll} 
 0 & 0 & 0 & -1 \\ 0 & 0 & 1 & 0 \\ 0 &
      1 & 0 & 0 \\ -1 & 0 & 0 & 0 \\  
\end{array}\right)
&=& \mbox{i}&\left(
\begin{array}{llll}  0 & -\sigma_2 \\ \sigma_2 & 0
\end{array}\right)\\
  \gamma_3 &=& 
\left(
\begin{array}{llll} 
0 & 0 & -\mbox{i} & 0 \\ 0 & 0 & 0 &
\mbox{i} \\
      \mbox{i} & 0 & 0 & 0 \\ 0 & -\mbox{i} & 0 & 0 \\  
\end{array}\right)
&=& \mbox{i}&\left(\begin{array}{llll} 
 0 & -\sigma_3 \\ \sigma_3 & 0
\end{array}
\right)\\
  \gamma_4 &=& \left(
\begin{array}{llll} 
0 & 0 & 1 & 0 \\ 0 & 0 & 0 & 1 \\ 1 & 0 &
      0 & 0 \\ 0 & 1 & 0 & 0 \\ 
\end{array}\right)
&=& &\left(\begin{array}{llll} 0 & \mathbf 1 \\ \mathbf 1 & 0
\end{array}\right)\\
  \gamma_5 &=& \left(
\begin{array}{llll}  1 & 0 & 0 & 0 \\ 0 & 1 & 0 & 0 
      \\ 0 & 0 & -1 & 0 \\ 0 & 0 & 0 & -1 \\ 
\end{array}\right)
\end{array}
\end{equation}

\end{document}